\RequirePackage[2020-02-02]{latexrelease}
\documentclass[%
reprint,
superscriptaddress,
twocolumn,
showpacs,
amsmath,amssymb,amsthm,amsfonts, 
aps,
prb,
floatfix,notitlepage,latexsym
]{revtex4-2}

\usepackage{graphicx}
\usepackage{comment,xcolor}

\usepackage{bm}
\usepackage{braket}
\usepackage{comment}

\newcommand{\lsec}[1]{\textit{#1.---}}

\newcommand{\tr}{\mathrm{tr}}
\newcommand{\sgn}{\mathrm{sgn}}

\newcommand{\cyc}{{k}}

\begin{document}
\title{Fermi's golden rule for heating in strongly driven Floquet systems}
\author{Tatsuhiko N. Ikeda}
\affiliation{Institute for Solid State Physics, University of Tokyo, Kashiwa, Chiba 277-8581, Japan}
\author{Anatoli Polkovnikov}
\affiliation{Department of Physics, Boston University, Boston, Massachusetts 02215, USA}
\date{\today}

\begin{abstract}
We study heating dynamics in isolated quantum many-body systems driven periodically at high frequency and large amplitude.
Combining the high-frequency expansion for the Floquet Hamiltonian with Fermi's golden rule (FGR), we develop a master equation termed the Floquet FGR.
Unlike the conventional one, the Floquet FGR correctly describes heating dynamics, including the prethermalization regime, even for strong drives, under which the Floquet Hamiltonian is significantly dressed, and nontrivial Floquet engineering is present.
The Floquet FGR depends on system size only weakly, enabling us to analyze the thermodynamic limit with small-system calculations.
Our results also indicate that, during heating, the system approximately stays in the thermal state for the Floquet Hamiltonian with a gradually rising temperature.
\end{abstract}
\maketitle

\section{Introduction}
Floquet engineering, i.e., controlling functionalities of physical systems by external periodic drives, has attracted renewed attention partly due to the advancement of laser technology~\cite{Goldman2014,Bukov2015,Holthaus2015,Eckardt2017,Oka2019}.
A key concept is the (approximate) local Floquet Hamiltonian $H_F$~\cite{Bukov2015} that is generically only defined when the driving period $T$ is short compared to the system's characteristic response time. Then $H_F$ approximately reproduces the actual unitary dynamics at short and intermediate times.
By cleverly designing the driving protocol, one can convert a commonplace undriven Hamiltonian $H_0$ into the ``dressed'' $H_F$ corresponding to intriguing physical systems such as topological insulators~\cite{Oka2009,Kitagawa2010,Perez-Gonzalez2019} and Floquet time crystals~\cite{Else2016,VonKeyserlingk2016,Yao2017}, part of which have been realized experimentally~\cite{Wang2013,Rechtsman2013,Jotzu2014,Choi2017,Zhang2017}.

Despite its usefulness, the local Floquet Hamiltonian $H_F$ fails to describe long-time dynamics in interacting/nonlinear systems. This is because the unitary evolution by $H_F$ introduces a new Floquet energy conservation law absent in the original driven model and thus cannot capture heating, which accompanies periodic driving eventually bringing isolated systems to the featureless infinite-temperature state~\cite{Lazarides2014,DAlessio2014,Kim2014}.
Nonetheless, especially for high-frequency drives, the heating has been rigorously proven to occur exponentially slowly in the driving frequency~\cite{Abanin2015,Mori2016,Kuwahara2016,Abanin2017}. In one-dimensional systems, a tighter, faster than exponential, bound was later found~\cite{Avdoshkin2020}.
At short and intermediate times, the dynamics is well approximated by the unitary evolution by $H_F$. While these rigorous results provide upper bounds, it is still elusive to obtain the heating rates accurately in concrete systems.
This especially applies to large-amplitude high-frequency drivings, where $H_F$ can significantly differ from the time-average Hamiltonian~\cite{Bukov2015}.

Several theoretical works have attempted to understand heating dynamics quantitatively.
Since heating is inherent to large systems, simulating those is numerically demanding and has become accessible only recently~\cite{Bukov2016heating,Machado2019,Machado2020,Luitz2020,Pizzi2020,Ye2020,Yin2021,Fleckenstein2021a,Fleckenstein2021b}.
Beyond simulations, a theory based on Fermi's golden rule (FGR) has been developed to describe heating rates~\cite{Mallayya2019,Mallayya2019x,Rakcheev2020}.
Yet, this theory is limited to weak drives, where $H_F{\approx}H_0$ up to small $1/\omega$ corrections. When strong, the driving is responsible for both heating and dressing $(H_0\to{H_F})$, it is quite subtle to separate the two effects, and the simple FGR does not work well~\cite{Fleckenstein2021a,Fleckenstein2021b} (see also Refs.~\cite{Zhao2021,Mori2021} for a random driving).

In this work, we develop a Floquet-theoretical extension of FGR (Floquet FGR) that can describe heating dynamics even in the strong-drive regime, where nontrivial Floquet engineering is possible with a well defined local Floquet Hamiltonian $H_F{\not\approx}H_0$.
The Floquet FGR calculation rapidly converges with the system size, enabling us to access the thermodynamic limit (TDL) more efficiently than the direct simulation.
We also show that, during heating dynamics, the system is thermal for $H_F$ at an instantaneous (inverse) temperature $\beta(t)$, which evolves self-consistently with the injected energy.
This result implies that the Floquet-Hamiltonian description survives even after heating sets in, where the finite-temperature Floquet engineering is realized.

\section{Floquet FGR}
We consider a periodically-driven quantum many-body system whose Hilbert-space dimension $d$ is large but finite.
Its time evolution is characterized by a Floquet unitary $U$, under which the stroboscopic evolution of a many-body (pure) state is given by $\ket{\Psi_\cyc}=U^\cyc\ket{\Psi_0}$ with $\ket{\Psi_0}$ being the initial state.

An approximate Floquet Hamiltonian $H_F$ plays a central role in understanding the periodically-driven system, and the associated approximate Floquet unitary $U_F$ is defined by
\begin{align}
U_F{\equiv}e^{-iH_F T}, 	
\end{align}
where $T$ is the driving period (and $\omega\equiv2\pi/T$).
There are various ways to define $H_F$, and the following argument equally applies to them.
For concreteness, we focus on the high-frequency (i.e., short-period) Magnus expansion~\cite{Magnus1954,Blanes2009} which gives $H_F$ as a power series for $T$.
For a stepwise drive (see Eqs.~\eqref{eq:spinH} and \eqref{eq:FloquetU} below), this expansion corresponds to the Baker-Campbell-Hausdorff (BCH) formula, giving $H_F$ by nested commutators between $H_0$ and the driving term $V$ accompanied by $T^N$ ($N=1,2,\dots$).
Thus the dressing strength $H_F-H_0$ roughly increases with $T\|V\|\sim\|V\|/\omega$.

The Magnus expansion is an asymptotic series and is generally not convergent. However, it was rigorously proven~\cite{Kuwahara2016} and confirmed numerically~\cite{Howell2019} that the expansion up to an optimal order $N_*=O(\omega)$ tends to converge to a local $H_F$ and well approximates the Floquet unitary: $U_F\approx{U}$.
If we continued the expansion still higher orders than $N_*$, the Magnus expansion would tend to diverge or, if converge in finite-size systems, approach a highly nonlocal Hamiltonian.
Throughout this work, we only consider low-order expansions, in which $H_F$ is local and $U_F\approx U$.
To the authors' knowledge, finding the optimal order $N_*$ is still an open question although reasonable estimates are given in inequality analyses~\cite{Kuwahara2016,Abanin2017}. Since $U_F\neq{U}$, the difference between the exact evolution $\ket{\Psi_k}$ and approximate one $\ket{\Psi_k^\mathrm{app}}=U_F^k\ket{\Psi_0}$ grows with $k$. This difference underlies the origin of Floquet heating. While previous works gave upper bounds on the heating rates~\cite{Abanin2015,Kuwahara2016,Avdoshkin2020}, they usually do not allow for quantitative estimates of these rates, which is the main subject of this paper.

Our idea of describing the heating dynamics is to focus on transitions between the eigenstates of $H_F$.
As $U\approx{U_F}$ and $[U_F,H_F]=0$, these states are approximately stationary under the actual evolution $U$, while residual transitions between them are caused by
\begin{align}\label{eq:dU}
\delta U\equiv U_F^\dag U\neq I,
\end{align}
where $I$ denotes the identity operator.
If $U_F=U$ held, $\delta{U}=I$ and no transitions could occur. Let us comment that there is an ambiguity in defining $\delta U$. Other possible choices are $\delta U=U U_F^\dagger$ or $\delta U=U_F^{\frac{1}{2}\dag}UU_F^{\frac{1}{2}\dag}$. We will find that these choices give the same physical result (see Eq.~\eqref{eq:wmn} below) and that the choice~\eqref{eq:dU} is most natural (see Appendix~\ref{app:symmw} for details). We assume that $H_F$ describes well the system dynamics within one period implying that $\delta U$ is close to identity for local bounded Hamiltonians~\cite{Kuwahara2016}).

More precisely, we assume that at each moment of time the system is well described by a local Floquet diagonal ensemble: $\rho(t)=\sum_nP_n(t)\ket{n}\bra{n}$, where $\ket{n}$ are the eigenstates of the Floquet Hamiltonian: $H_F\ket{n}=E_n\ket{n}$ $(n=1,2,\dots,d)$, and correspondingly $U_F\ket{n}=e^{-i\theta_n}\ket{n}$ ($\theta_n{\equiv}E_nT$). 
The weights $P_n(t)$ are determined by the master equation: 
\begin{align}\label{eq:master}
\frac{dP_n(t)}{dt}=\sum_m\left[ w_{m\to n}P_m(t)-w_{n\to m}P_n(t)\right],	
\end{align}
where $w_{m{\to}n}$ are the transition rates from $\ket{m}$ to $\ket{n}$ given by (see below for its derivation):
\begin{align}\label{eq:wmn}
w_{m\to n}=\omega\sum_{l\in\mathbb{Z}}\delta(\theta_n-\theta_m-2\pi l)|\braket{n|\delta U|m}|^2
\end{align}
for $m\neq n$.
Clearly, we can freely assign the diagonal elements $w_{n{\to}n}$ without changing the master equation~\eqref{eq:master}. Note that in isolated systems generally the only attractor of the master equation is an infinite temperature state~\cite{DAlessio2016}, i.e. $P_n={\rm const}$. We term this formalism as the Floquet FGR since it inherits the spirit of original FGR: The transition rate $w_{m{\to}n}$ is time-independent and determined by the perturbation matrix elements $\braket{n|{\delta}U|m}$.

The Floquet FGR is a natural extension of the standard (bare) FGR~\cite{Mallayya2019} for Hamiltonians of type $H(t)=H_0+V(t)$.
In the standard FGR, we consider the energy eigenstates of the undriven Hamiltonian ($H_0\ket{n}_0=E_n^{(0)}\ket{n}_0$) and the transitions between them when the driving amplitude $\|V(t)\|$ is small.
The two FGRs coincide when the driving amplitude is small and $H_F\approx{H_0}$ (see Appendix~\ref{app:proof} for a proof).

Before closing this section, let us formally derive Eq.~\eqref{eq:wmn}. The central idea of the derivation is an assumption that $\delta U=U_{F}^{\dagger} U$ is a unitary operator, which is close to the identity $I$. We can then write $U$ as
\begin{align}
	U=U_{F} U_{F}^{\dagger}U = U_{F}\delta U \approx U_F(I+i\delta K),
\end{align}
where $\delta K$ is some ``small'' Hermitian operator. If e.g. $U_F$ is obtained using the high frequency expansion then $\delta K$ is suppressed correspondingly to the order of the expansion power of the driving frequency. Note that have $\braket{n|\delta U|m}\approx i \braket{n|\delta K|m}$ for $m\neq n$.
We consider the transition probability $p_{m \rightarrow n}$ from $\ket{m}$ to $\ket{n}$ ($m\neq n$) during $N$ $(\gg1)$ Floquet cycles. In the  leading-order approximation in $\delta K$ it is given by
\begin{multline}\label{seq:pmn}
	p_{m \rightarrow n} \equiv\left| \left\langle n\left|U^{N}\right| m\right\rangle\right|^{2} 
\approx \left|  \sum_{k=1}^N\langle n|U_{F}^{k} \delta K U_{F}^{N-k}| m\rangle \right|^{2} \\
\approx \left|\sum_{k=1}^N \mathrm{e}^{i\left(\theta_{n}-\theta_{m}\right) k}\right|^2 |\langle n|\delta U| m\rangle|^{2}.
\end{multline}
Now, we note
$\sum_{k=1}^N \mathrm{e}^{i\left(\theta_{n}-\theta_{m}\right) k}\approx 2\pi \sum_{l=-\infty}^\infty \delta(\theta_n-\theta_m -2\pi l)$,
which leads to
\begin{align}\label{seq:sqphase}
	\left|\sum_{k=1}^N \mathrm{e}^{i\left(\theta_{n}-\theta_{m}\right) k}\right|^2 \approx 2\pi N \sum_{l\in\mathbb{Z}} \delta(\theta_n-\theta_m -2\pi l),
\end{align}
where the factor of $N$ comes from the observation that for $\theta_n-\theta_m=2\pi \ell$ we have $\sum_{k=1}^N \mathrm{e}^{i\left(\theta_{n}-\theta_{m}\right) k}=\sum_{k=1}^N 1=N$. Combining Eqs.~\eqref{seq:pmn} and\eqref{seq:sqphase} and defining the transition probability per unit time $w_{m\to n}=p_{m\to n}/(NT)=p_{m\to n} \omega/(2\pi N)$ we derive Eq.~\eqref{eq:wmn}.

\section{Self-consistent evolution of temperature}
The master equation~\eqref{eq:master} together with the expression for the transition rates~\eqref{eq:wmn} is sufficient for computing heating of Floquet systems.
To advance further, we will assume, from now on, that $H_F$ is quantum ergodic and the eigenstate thermalization hypothesis~\cite{Deutsch1991,Srednicki1994,Rigol2008} holds.
In such a case, aside from initial transients, we can regard $\ket{\Psi_\cyc}$ as thermal at an appropriate (inverse) temperature $\beta$~\cite{Polkovnikov2011,Eisert2015,DAlessio2016,Mori2018}.
Thus, at each time $t$, it is reasonable to set the canonical distribution
\begin{align}\label{eq:Pn}
\rho^F_{\beta(t)}=\sum_n P_n^{\beta(t)}\ket{n}\bra{n},\quad
P_n^{\beta(t)}\equiv e^{-\beta(t)E_n}/Z_t	
\end{align}
with the time-dependent temperature $\beta(t)$ and partition function $Z_t=\sum_n e^{-\beta(t)E_n}$.
Within this ansatz, expectation values of observables $O$ are obtained as $\langle{O}\rangle_t=\sum_n\braket{n|O|n}P_n^{\beta(t)}$.
In driven isolated systems, generally, the correct statistical ensemble is different from the Gibbs ensemble~\cite{Bunin2011}, but the emerging differences vanish in the TDL.

We then can obtain a self-consistent evolution equation for $\beta(t)$ by first computing the Floquet energy $E_F$ and its variance $\sigma_F^2$ corresponding to a given temperature $\beta(t)$:
\begin{align}\label{eq:EF}
	 E_F = \sum_n E_n P_n^{\beta(t)},  \quad
	 \sigma_F^2=\sum_n E_n^2 P_n^{\beta(t)}-E_F^2.
\end{align}
Then, using the chain rule together with Eq.~\eqref{eq:master}, we find
\begin{equation}\label{eq:SdotP}
	\frac{d\beta}{dt}=-\frac{1}{\sigma_F^2}\frac{dE_F}{dt},\quad \frac{dE_F}{dt}=\sum_{n,m} P_{n}^{\beta(t)} (E_n-E_m) w_{n\to m},
\end{equation}
where we used the fact that for a canonical ensemble: $\sigma_F^2=-dE_F/d\beta$~\cite{Kardar2007}.

\begin{figure}[tb]
	\includegraphics[width=\columnwidth]{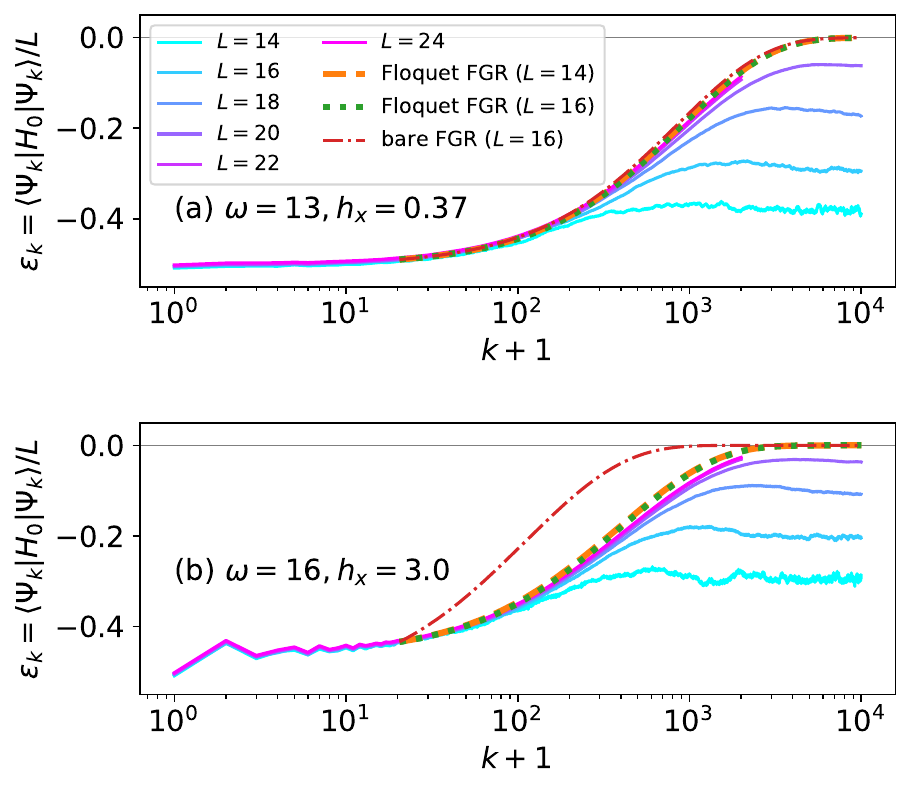}
	\caption{
	Typical heating dynamics for $\epsilon_\cyc=\braket{\Psi_\cyc|H_0|\Psi_\cyc}/L$ under (a) weak ($\omega=13$ and $h_x=0.37$) and (b) strong ($\omega=16$ and $h_x=3.0$) drives.
	The horizontal axis denotes the Floquet cycle $k$.
	Solid curves are the results obtained by numerically simulating the unitary evolution.
	Dashed (dotted) curves show that obtained by the Floquet FGR with the 6th-order Magnus expansion at $L=14$ ($L=16$), and dash-dotted ones by the bare FGR at $L=16$.
	}
	\label{fig:dynamics}
\end{figure}

\section{Numerical verification}
Let us verify how well the Floquet FGR describes actual heating dynamics in concrete models.
Following Ref.~\cite{Machado2019},
we consider a periodically-driven spin-chain Hamiltonian $H(t)=H_0+g(t)V$ with
\begin{multline}\label{eq:spinH}
H_0 = J\sum_{i} \sigma_{i}^{z} \sigma_{i+1}^{z}+ J' \sum_{i} \sigma_{i}^{z} \sigma_{i+2}^{z}+h_z \sum_{i} \sigma_{i}^{z}\\+ J_{x} \sum_{i} \sigma_{i}^{x} \sigma_{i+1}^{x},\quad V=h_x \sum_i \sigma_i^x,
\end{multline}
and $g(t)=\sgn(\cos(\omega t))$, which give the Floquet unitary
\begin{equation}\label{eq:FloquetU}
U=e^{-i(H_0+V)T/4}e^{-i(H_0-V)T/2}e^{-i(H_0+V)T/4}.	
\end{equation}
Here, $\sigma_i^\alpha$'s are the Pauli matrices acting on site $i$ $(=1,2,\dots,L)$, and the periodic boundary conditions are imposed.
We set $J=-1$, $h_z=0.6$, $J'=-0.4$, and $J_x=0.75$, ensuring that $H_0$ is ergodic enough.
Taking a thermal pure state~\cite{Sugiura2012,Sugiura2013} at energy density $\epsilon_0$ as our initial state $\ket{\Psi_0}$ (see Appendix~\ref{sec:initstate} for details), we numerically obtain $\ket{\Psi_\cyc}=U^\cyc\ket{\Psi_0}$ by the Krylov evolution method~\cite{Moler2003}.
Below, we focus on the high-frequency drives $\omega\gtrsim10\gg{J,J'}$.
In the other regime $\omega\lesssim 10$, the heating occurs very fast, and there is no well-defined Floquet Hamiltonian.

Typical heating dynamics of the energy density $\epsilon_\cyc\equiv\braket{\Psi_\cyc|H_0|\Psi_\cyc}/L$ is shown in Fig.~\ref{fig:dynamics} for two parameter sets, $(\omega,h_x)=(13,0.37)$ and $(16,3.0)$.
In both cases, strong finite-size effects make heating saturate at some energy density, and the infinite-temperature state $(\epsilon_\cyc=0)$ can be reached only at large system sizes.
Although the computational complexity limits us to $L\le24$, we observe systematic tendencies to the TDL. 

For weak driving amplitudes, the Floquet and bare FGRs give similar results, both describing the heating dynamics well, as is evident from Fig.~\ref{fig:dynamics}(a).
The bare FGR's accuracy is also consistent with Ref.~\cite{Mallayya2019}.
Note that, to avoid the initial transients, we have evolved $\beta(t)$ from $t=20T$, at which $\beta$ is chosen to satisfy  $\text{tr}[\rho^F_{\beta(t)}H_0]/L=\epsilon_{\cyc=20}$, where $\rho^F_{\beta(t)}$ is given by Eq.~\eqref{eq:Pn}.
For the Floquet FGR, we have used $H_F$ obtained by the 6th-order Magnus expansion (we will discuss the order dependence below) and approximated the delta function in Eq.~\eqref{eq:wmn} by a Gaussian of width $\delta\theta=T{\delta}E$ with ${\delta}E=0.03L$.
As Fig.~\ref{fig:dynamics}(a) shows, the FGRs capture the unitary dynamics in the TDL reasonably well. Remarkably, the FGR shows a rapid size convergence as $L$ increases giving a clear advantage of FGR in studying the TDL. In the following we will use $L=14$ for FGR calculations.

For strong drivings, the Floquet FGR better describes dynamics than the bare one, as shown in Fig.~\ref{fig:dynamics}(b).
This happens because the bare FGR heating rates are proportional to $h_x^2$, obviously overestimating them at large $h_x$.
Meanwhile, in the Floquet FGR, both $H_F$ and $\delta{U}$ depend intricately on $h_x$ capturing well the actual heating dynamics.
The Floquet FGR's accuracy justifies the assumption that, at each instance of time, the system is in approximate equilibrium for the dressed $H_F$.

\begin{figure}[tb]
	\includegraphics[width=\columnwidth]{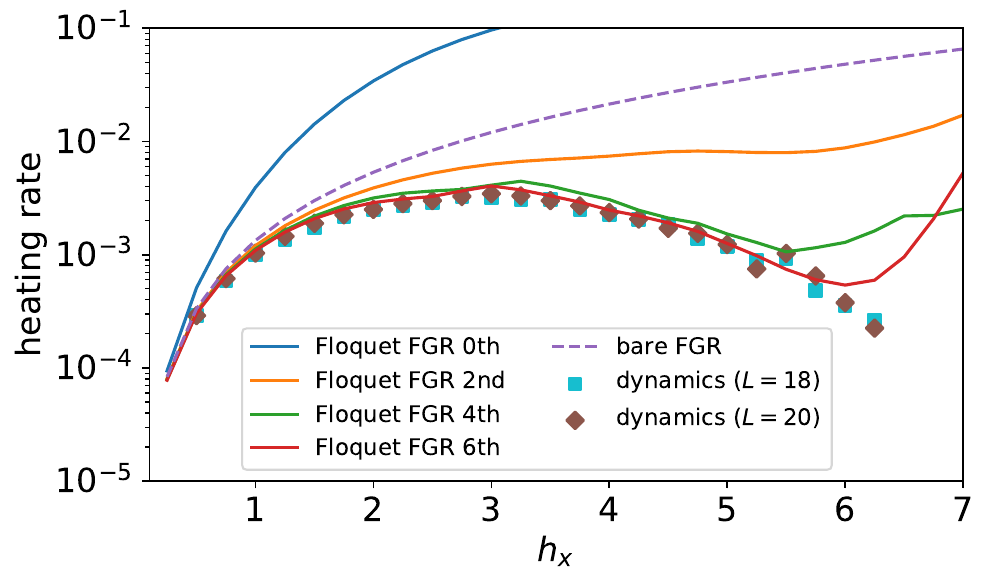}
	\caption{
	Heating rates extracted from unitary dynamics at energy density $\epsilon=-0.48$ by the linear least squares.
	The solid (dashed) curves show the heating rates obtained by the Floquet (bare) FGR, and the Floquet FGR's order from 0th to 6th in the legend shows that of the Magnus expansion for $H_F$ (see also text).
	All the FGR calculations are for $L=14$.
	}
	\label{fig:heating_rate}
\end{figure}

\section{Suppression of heating at strong drives}
To uncover further nontrivial aspects of strong drives, we systematically study how the heating rate depends on the driving amplitude $h_x$.
With $\omega=16$ fixed, we simulate the unitary evolution for various $h_x$ and extract the heating rates at an energy density $\epsilon_k=-0.48$ as the slope $d\epsilon/dt$ obtained by the linear least-squares fit (see Appendix~\ref{app:least-square} for detail).
This energy density is chosen to minimize finite size effects for unitary evolution simulations, which are especially strong near the saturation energy as seen in Fig.~\ref{fig:dynamics} for $k\gtrsim10^3$.
The heating rates thus extracted from dynamics and from FGRs are shown in Fig.~\ref{fig:heating_rate}.
Interestingly, the rate does not monotonically increase with $h_x$ but becomes maximum at $h_x\approx3$ then decreases for further larger $h_x$.
This nonmonotonic behavior implies the Floquet prethermalization (see also Appendix~\ref{app:least-square}) and is consistent with earlier results in similar models~\cite{Das2010,Haldar2018}.

The nonmonotonic heating rate in Fig.~\ref{fig:heating_rate} is well captured by the Floquet FGR with $H_F$ obtained by the Magnus expansion at sufficiently high order.
The Floquet-FGR results are obtained by the following three steps.
First, we use the Magnus expansion of the  Floquet unitary $U$ to compute the Floquet Hamiltonian as a power series of $T$. Note that for symmetric drives only odd terms show up in the expansion~\cite{Bukov2015}. Second, we find the temperature $\beta$ that reproduces our target energy density $\epsilon=-0.48=\text{tr}(\rho^F_{\beta}H_0)/L$.
Note that $\epsilon$ is the physical energy of the system. And finally, using Eqs.~\eqref{eq:wmn} and \eqref{eq:SdotP}, we compute the temperature increase $d\beta$ in a small time step $dt$ and the corresponding new energy $\epsilon'$. Thus, we obtain the heating rate as $(\epsilon'-\epsilon)/dt$.

We remark that the Floquet FGR, even at the 6th order, fails to describe the heating rate for very large amplitudes $h_x\gtrsim6$ (see Fig.~\ref{fig:heating_rate}).
This is due to the failure of the Magnus expansion for $H_F$ as it contains terms of the order $O(T^{N}h_x^{N+1})$, which do not decrease with $N$.
Applying the Floquet FGR in this limit would require a different approach for finding the approximate Floquet Hamiltonian, for example, high-frequency expansion in the rotating frame~\cite{Bukov2015,Bukov2016sw,Haldar2021}, replica resummation of the BCH series~\cite{Vajna2018}, or flow equation method~\cite{Vogl2019,Claassen2021}.
We anticipate that the Floquet FGR will remain accurate if we use one of those methods.

\begin{figure}[tb]
	\includegraphics[width=\columnwidth]{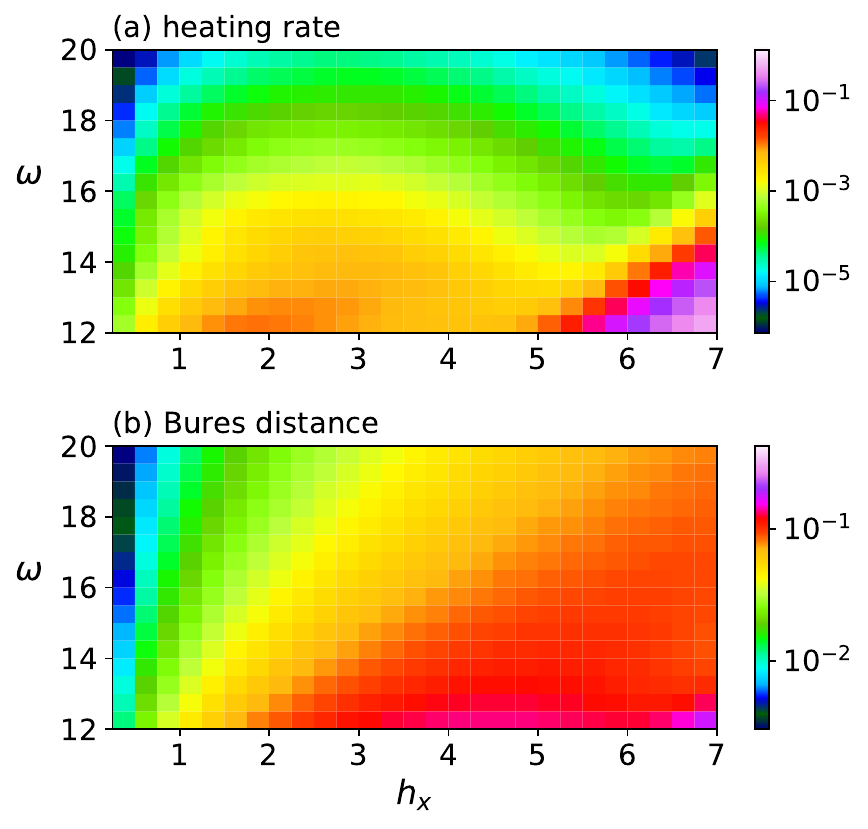}
	\caption{(a) Heating rate $d\epsilon/dt$ calculated by the Floquet FGR with the 6th-order $H_F$ at temperature $\beta=0.1$.
	(b) Bures distance $D_B(\rho_\mathrm{th}^F,\rho_\mathrm{th}^0)$~\eqref{eq:bures} between thermal states for $H_F$ and $H_0$ at temperature $\beta=0.1$. In each panel, we have used $L=14$.
	}
	\label{fig:heatmap}
\end{figure}

It is instructive to extend the previous analysis to the whole $(h_x,\omega)$-plane. The Floquet FGR enables this efficiently due to its weak system-size dependence. It is thus sufficient to use a relatively small system size $L=14$ to extract the two-dimensional heat map shown in Fig.~\ref{fig:heatmap}(a). Here different colors encode different heating rates $d\epsilon/dt$ calculated using the 6th-order Floquet FGR at a fixed temperature $\beta=0.1$. There is a clear triangular-shaped region in the bottom-right corner of Fig.~\ref{fig:heatmap}(a), where the Floquet FGR fails. This region precisely corresponds to the region of failure of the Magnus expansion. Also, the plot is cut below $\omega=12$ as, at lower frequencies, there is no Floquet prethermal phase, and the Floquet Hamiltonian becomes ill-defined.

Aside from the corner, the heating-rate map in Fig.~\ref{fig:heatmap}(a) is reliable and provides key information.
First, as $\omega$ increases with $h_x$ fixed, the heating rate is rapidly suppressed as expected~\cite{Abanin2015,Kuwahara2016,Avdoshkin2020}.
Second, the rate is a non-monotonic function of $h_x$ for all shown frequencies. Consequently, the heating rate is suppressed at either the top-left or top-right corner of Fig.~\ref{fig:heatmap}(a).

From the point of view of Floquet engineering, the right-top corner in Fig.~\ref{fig:heatmap}(a) is most interesting as there both the heating rate is suppressed and the Floquet Hamiltonian is strongly dressed, i.e., is significantly different from $H_0$. To quantify the difference between $H_F$ and $H_0$ we introduce the Bures distance between the corresponding thermal distributions~\cite{Bures1969,Helstrom1967}:
\begin{equation}\label{eq:bures}
D_B(\rho,\sigma)=\left[2(1-\tr\sqrt{\rho^{1/2}\sigma\rho^{1/2}}\right]^{1/2}	
\end{equation}
where $\rho=\rho_\beta^F$ and $\sigma=\rho_\beta^0$ are the thermal density matrices for $H_F$ and $H_0$, respectively, evaluated at the same temperature $\beta$.
Large Bures distance means that the Floquet prethermal state is significantly different from the thermal state of the undriven Hamiltonian $H_0$. Figure~\ref{fig:heatmap}(b) shows the Bures distance at $\beta=0.1$, indicating the expected tendency of strong renormalization of $H_F$ at large driving amplitudes. We remark that Fig.~\ref{fig:heatmap}(b) has clear linear structures separating different color regions. This reflects that the Magnus expansion is indeed dominated by the powers of the dimensionless ratio $h_x/\omega$.


\section{Other models and possible experiments}

Spin models analyzed above can be realized, for example, in trapped ions~\cite{Leibfried2003,Blatt2012}, Rydberg atoms~\cite{Ebadi2021}, nuclear spins~\cite{Peng2021}, nitrogen-vacancy centers~\cite{Lang2015}, superconducting qubits~\cite{Kjaergaard2020}, and other systems. In particular, heating in Floquet spin systems was observed in Refs.~\cite{Peng2021,Beatrez2021}. Even wider applications of the developed Floquet FGR can be anticipated in materials, where the recent advances of terahertz spectroscopy and pump-probe methods enable us to access time scales, where electron systems are effectively decoupled from phonons and thus can be treated as isolated~\cite{Ron2020,Liu2020}.

While analyzing specific experiments requires its own systematic analysis, which is beyond the scope of this work, we want to illustrate that the methodology developed here can be applied to interacting electrons described by a driven Hubbard model:
\begin{align}
H_0&=-t_1\sum_{i=1}^{L-1}\sum_{s=\uparrow,\downarrow}(c_{i,s}^{\dag}c_{i+1,s}+\mathrm{H.c.})\notag\\
&-t_2\sum_{i=1}^{L-2}\sum_{s=\uparrow,\downarrow}(c_{i,s}^{\dag}c_{i+2,s}+\mathrm{H.c.})
+U{\sum_{i=1}^L}n_{i,\uparrow}n_{i,\downarrow}
+H_\mathrm{sb},\notag\\
V&=E_0\sum_{i=1}^L\left(i-\frac{L+1}{2}\right)(n_{i,\uparrow}+n_{i,\downarrow}).
\end{align}
The undriven Hamiltonian $H_0$ is an ergodic Fermi-Hubbard chain of length $L$ that includes the next-nearest-neighbor hopping ($t_1=1$, $t_2=0.5$, and $U=4$) and small symmetry-breaking term $H_\mathrm{sb}=h_{b}(n_{1,\uparrow}-n_{1,\downarrow})+\mu_{b}(n_{L,\uparrow}+n_{L,\downarrow})$ with $h_b=\mu_b=10^{-3}$ on the chain boundaries~\cite{Mondaini2015}.
Our driving term $g(t)V$ is the alternating electric field along the chain represented by the linear electric potential $V$ with $E_0$ being the electric-field strength.
Like in the spin model, the total Hamiltonian is $H(t)=H_0+g(t)V$, where $g(t)=\sgn(\cos(\Omega t))$.
Following Ref.~\cite{Mondaini2015}, we consider even $L$'s and the 1/4-filling, for which $N_\uparrow+N_\downarrow=L/2$ and $N_\uparrow=N_\downarrow$ ($N_\downarrow+1$) when $L/4$ is integer (noninteger).
Their values and the corresponding Hilbert-space dimensions are shown in Table~\ref{table:dimH}.
We simulate the unitary evolution by the Krylov evolution method, and the initial state is created by the method in Sec.~\ref{sec:initstate}.
The FGR calculations have also been done similarly with the exact diagonalization of $H_F$ and the thermal ansatz with the time-dependent temperature $\beta(t)$.

\begin{table}
  \caption{System size, number of particles, and Hilbert-space dimension for the Fermi-Hubbard model at $1/4$-filling.}
  \label{table:dimH}
  \centering
  \begin{tabular}{lccc}
    \hline
    $L$  & $N_\uparrow$ & $N_\downarrow$  & $D$  \\
    \hline \hline
    8  &  2 & 2 & 784 \\
    10  & 3 & 2  & 5400 \\
    12 &  3 & 3 & 48400 \\
    14  & 4 & 3  &  3312400 \\
    \hline
  \end{tabular}
\end{table}

\begin{figure}[tb]
	\includegraphics[width=\columnwidth]{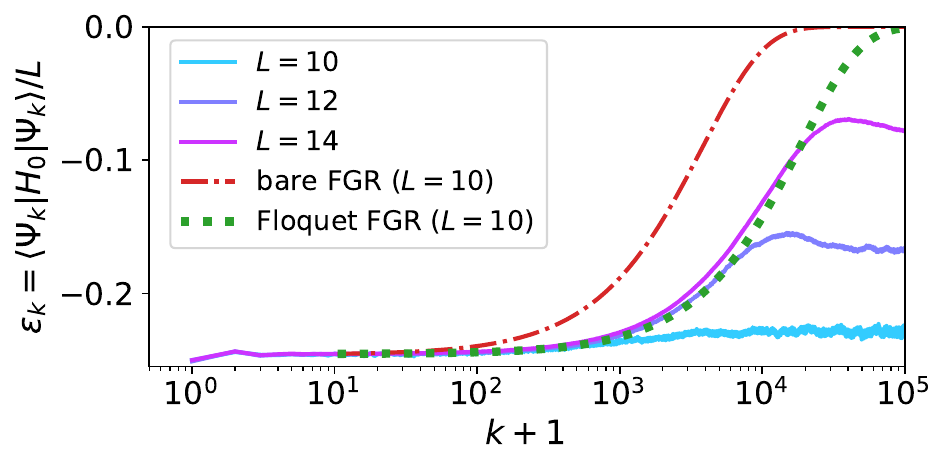}
	\caption{
	Typical heating dynamics in the open Fermi-Hubbard chain for a strong drive ($E_0=4$ and $\omega=10$).
	Solid curves are the results obtained by numerically simulating the unitary evolution.
	Dotted (dash-dotted) curves show that obtained by the 6th-order-Floquet (bare) FGR at $L=10$.}
	\label{fig:fermi_hubbard}
\end{figure}

Like the spin model, the Floquet FGR works better than the bare one for strong drives in the Fermi-Hubbard model.
Figure~\ref{fig:fermi_hubbard} shows a typical heating dynamics for such a case ($E_0=4$ and $\omega=10$) from an initial energy $\epsilon=-0.25$, compared with the bare and 6th-order-Floquet FGRs starting from $t=10T$.
This result is qualitatively similar to that in the spin model in Fig.~\ref{fig:dynamics}, evidencing the general applicability of the Floquet FGR and opening the possibility of experimental verification in Fermi-Hubbard models.
Thus, one could use those models simulated in ultracold atoms~\cite{Scherg2021} and possibly in solid-state materials if dissipation is negligible.

\section{Summary and prospect}
We formulated the Floquet FGR by combining the high-frequency expansion with the conventional Fermi's golden rule. We verified that it correctly describes heating dynamics of periodically-driven many-body quantum systems even in the strong-drive regime, where the bare FGR fails. The Floquet FGR is practically useful because it captures the TDL already at a small system sizes. High accuracy of the Floquet FGR indicates that, aside from initial transients, the system we analyzed can be characterized as an equilibrium state for the Floquet Hamiltonian $H_F$ with a slowly changing inverse temperature $\beta(t)$.

We leave some important generalizations for future work.
First, we have focused on ergodic $H_F$ having no relevant conserved quantity and utilized the thermal ansatz~\eqref{eq:Pn}.
Yet, this ansatz should be modified appropriately when $H_F$ has conserved quantities like in, e.g., integrable models~\cite{Gritsev2017,Haldar2021,Haldar2021b} and Floquet time crystals (or pi glasses)~\cite{Machado2020,Luitz2020,Pizzi2020}.
Second, it is interesting to combine this approach with other sophisticated schemes for finding approximate Floquet Hamiltonians.
Finally, we are planning to apply the Floquet FGR approach to open systems to find transient and steady states in driven-dissipative systems~\cite{Engelhardt2019,Ikeda2020}.

\lsec{Note added}
The terminology, Floquet FGR, is also used for slightly different concepts in kinetics~\cite{Genske2015,Knap2016,Weidinger2017,Rudner2020}.

\section*{Acknowledgements}
Numerical exact diagonalization in this work has been performed with the help of the QuSpin package~\cite{Weinberg2017,Weinberg2019}.
The computation in this work has been done using the facilities of
the Supercomputer Center, the Institute for Solid State Physics, the
University of Tokyo.
T. N. I. was supported by JSPS KAKENHI Grant No. JP21K13852.
A. P. was supported by NSF under Grant No. DMR-1813499 and DMR-2103658 and by the AFOSR under Grant No. FA9550-16-1-0334 and FA9550-21-1-0342.

\appendix

\section{Symmetry of transition rates}\label{app:symmw}
We show that $w_{m\to n}$ is approximately symmetric: $w_{m\to n}\approx w_{n\to m}$.
For this, let us use the $w_{m\to n}$'s expression by $\delta U$ in Eq.~\eqref{eq:wmn}.
As $U_F^\dag U$ is a unitary matrix, there exists a Hermitian matrix $\delta K$ such that $U_F^\dag U=e^{i\delta K}$.
Suppose that $H_F$ is a good Floquet Hamiltonian and $U_F\approx U$, meaning that $\braket{n|\delta U|m}\approx i\braket{n|\delta K|m}$ for $m\neq n$ and hence
\begin{align}\label{seq:appwmn}
	w_{m\to n} \approx \omega \sum_{l\in\mathbb{Z}} \delta(\theta_n-\theta_m-2\pi l)|\langle n|\delta K|m\rangle |^2.
\end{align}
Here we note that $|\langle n|\delta K|m\rangle |$ is symmetric under interchange of $m$ and $n$ as $\delta K$ is Hermitian and that $\sum_{l\in\mathbb{Z}} \delta(\theta_n-\theta_m-2\pi l)$ is also symmetric as $\delta(x)$ is an even function.
Therefore, the approximate $w_{m\to n}$ in Eq.~\eqref{seq:appwmn} is symmetric $w_{m\to n}=w_{n\to m}$.

We remark that, when $\braket{n|U|m}$ is a symmetric matrix, $w_{m\to n}=w_{n\to m}$ holds exactly.
To show this, we use $\braket{n|\delta U|m}=\braket{n|U_F^\dag U|m}=e^{i\theta_n}\braket{n|U|m}$, having
\begin{equation}
    	w_{m\to n}= \omega \sum_{l\in\mathbb{Z}} \delta(\theta_n-\theta_m-2\pi l)|\langle n|U|m\rangle |^2.
\end{equation}
This expression implies that, if $\braket{n|U|m}$ is a symmetric matrix, $w_{m\to n}$ is also symmetric.

As one can check easily, $\braket{n|U|m}$ is symmetric in the example model analyzed in the main text.
More generally, $\braket{n|U|m}$ is symmetric if there exists an appropriate basis where the $H_F$ is real symmetric, and the driving protocol is time-reversal symmetric $H(T-t)={}^t\! H(t)$.

\section{Equivalence of bare and Floquet FGRs for small driving amplitudes}\label{app:proof}
Let us consider the following periodic Hamiltonian
\begin{align}
	H(t)&=H_0+g(t)V,\notag\\
	g(t)&=\sum_{l\in\mathbb{Z}}g_l e^{-il\omega t},\notag\\
	g_l &=\int_0^T\frac{ds}{T} g(s)e^{il \omega s},\label{seq:hamg}
\end{align}
where $g(t+T)=g(t)$ and the driving amplitude $|g(t)|$ is small.
The bare FGR~\cite{Mallayya2019} is based on $H_0$'s eigenbasis $(H_0\ket{n}_0=E_n^{(0)}\ket{n}_0)$ and gives the transition rate from $\ket{m}_0$ to $\ket{n}_0$ as
\begin{align}\label{seq:wbare}
	w_{m\to n}^\mathrm{bare\,FGR} = 2\pi\sum_{l\in\mathbb{Z}}|V_{nm}|^2\ |g_l|^2 \delta(E_{nm}^{(0)}-l\omega)
\end{align}
with $V_{nm}\equiv {}_0\!\bra{n}V\ket{m}_0$ and $E_{nm}^{(0)}\equiv E_n^{(0)}-E_m^{(0)}$.

Our aim here is to prove the equivalence of bare and Floquet FGRs in the limit of $g\to0$ ($g$ symbolically stands for the driving amplitude).
Specifically, we prove that Eq.~\eqref{seq:wbare} coincides with Eq.~\eqref{eq:wmn} in the leading order of $g$ as $g\to0$.
To prove this, we begin by noticing that $H_F=H_0+O(g)$, meaning that $\ket{m}\to\ket{m}_0$ and $E_m\to E_m^{(0)}$ in this limit.
Thus, we can approximate Eq.~\eqref{eq:wmn} as
\begin{align}
	w_{m\to n} &\approx \frac{2\pi}{T}\sum_{l\in\mathbb{Z}} \delta(E_n^{(0)}T-E_m^{(0)}T-2\pi l)|{}_0\!\langle n|\delta U |m\rangle_0 |^2\\
	&=\frac{2\pi}{T^2}\sum_{l\in\mathbb{Z}} \delta(E_{nm}^{(0)}-l\omega)|{}_0\!\langle n|\delta U |m\rangle_0 |^2,\label{seq:wmn_smallg}
\end{align}
where we have used $\delta(ax)=\delta(x)/|a|$.

To obtain equivalence between Eqs.~\eqref{seq:wmn_smallg} and \eqref{seq:wbare}, we look into the matrix elements ${}_0\!\langle n|\delta U |m\rangle_0$.
To do this, we recall that
\begin{align}\label{seq:dU}
	\delta U = U_F^\dag U  \approx e^{iH_0T}U
\end{align}
and notice that $e^{iH_0T}U$ is the one-cycle evolution in the interaction picture
\begin{align}\label{seq:intpic}
	e^{iH_0T}U  = \exp_+\left[ -i\int_0^T ds\, g(s) V_I(s) \right]
\end{align}
with $V_I(s)\equiv e^{iH_0s}Ve^{-iH_0s}$ and $\exp_+$ denotes the time-ordered exponential.
Substituting Eq.~\eqref{seq:intpic} into Eq.~\eqref{seq:dU} and taking its matrix element, we have, for $m\neq n$,
\begin{align}
	{}_0\!\langle n|\delta U |m\rangle_0
	&\approx -i\int_0^Tds\,g(s) {}_0\!\langle n|V_I(s) |m\rangle_0 \notag\\
	&= -iV_{nm}\int_0^Tds\, g(s)e^{iE_{nm}^{(0)}s}.\label{seq:dUel}
\end{align}
Finally, we substitute Eq.~\eqref{seq:dUel} into Eq.~\eqref{seq:wmn_smallg} and invoke $\delta(E_{nm}^{(0)}-l\omega)$ that allows us to replace $E_{nm}^{(0)}$ by $l\omega$, obtaining
\begin{align}
	w_{m\to n}
	&=\frac{2\pi}{T^2} \sum_{l\in\mathbb{Z}} \delta(E_{nm}^{(0)}-l\omega)|V_{nm}|^2 \left| \int_0^Tds\, g(s)e^{il\omega s} \right|^2\notag\\
	&= 2\pi\sum_{l\in\mathbb{Z}}|V_{nm}|^2\ |g_l|^2 \delta(E_{nm}^{(0)}-l\omega),\label{seq:wFloquetSmallg}
\end{align}
where we have used the definition of $g_l$ in Eq.~\eqref{seq:hamg}.
Equations~\eqref{seq:wbare} and \eqref{seq:wFloquetSmallg} mean the equivalence between bare and Floquet FGRs in the limit of small driving amplitude $g\to0$.

\section{Initial state preparation}\label{sec:initstate}
In the main text, we have stated that $\ket{\Psi_0}$ is at the energy density $\epsilon_0$.
The explicit procedure to generate $\ket{\Psi_0}$ is what was developed in Ref.~\cite{Sugiura2012}.
First, we take $\ket{\psi_0}$ randomly following the uniform Haar measure on the whole Hilbert space of dimension $2^L$.
Second, we make iteratively
\begin{align}	
\left|\psi_{p+1}\right\rangle & \equiv \frac{(l-\hat{h})\left|\psi_{p}\right\rangle}{ \|(l-\hat{h})\left|\psi_{p}\right\rangle \|},\qquad
u_{p} \equiv\langle\psi_{p}|\hat{h}| \psi_{p}\rangle,
\end{align}
where $\hat{h}\equiv H_0/L$ and we have set $l=50$.
As shown in Ref.~\cite{Sugiura2012}, the energy density $u_p$ gradually decreases as we proceed, and we encounter $p_0$ such that $u_{p_0+1}>\epsilon_0 \ge u_{p_0}$.
Then, we set $\ket{\Psi_0}=\ket{\psi_{p_0}}$.
All the presented results are obtained for a single realization of $\ket{\psi_{p_0}}$, and we have confirmed that average over realizations does not change the results very much especially for larger system sizes like $L=20$.

\section{Heating rate extraction}\label{app:least-square}

\begin{figure*}[tb]
\begin{center}
	\includegraphics[width=18cm]{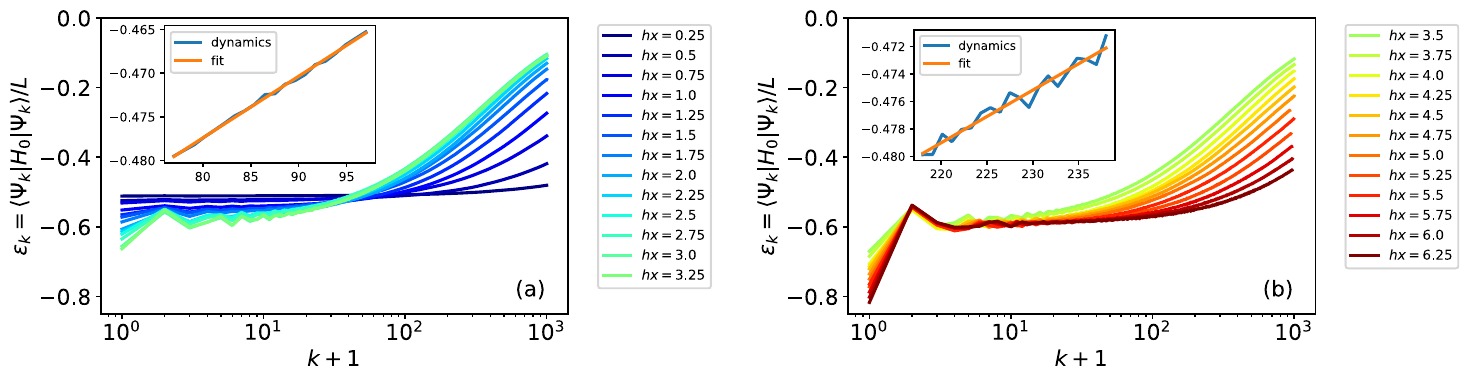}
	\caption{Heating dynamics for various driving amplitude $h_x$ in (a) $0.25\le h_x \le 3.25$ and (b) $3.5\le h_x \le 6.25$ calculated at frequency $\omega=16$ and system size $L=20$.
	The inset shows the first 20 Floquet cycles in which $\epsilon_k>-0.48$ for $h_x=1.5$ in panel (a) and $h_x=5.5$ in panel (b).
	The inset also shows the linear least squares fit.
	}
	\label{sfig:heating}
\end{center}
\end{figure*}

We supplement technical details for extracting heating rates shown in Fig.~2 in the main text.
We simulate the unitary evolution for $h_x=0.25\times{j}$ ($j=1,2,\dots,25$) with $\omega=16$ being fixed.
To obtain the heating rate from exact simulations, we initialize the system at $\epsilon_0=-0.5-0.05h_x$ and extract the slope of the energy density $\epsilon_k$ when it reaches a fixed value $\epsilon_k=-0.48$ from the linear least-squares fit.
Figure~\ref{sfig:heating} shows the simulated heating dynamics for each $hx$ at $L=20$ with $\omega=16$ being fixed.

We remark that the Floquet prethermalization is manifest in Fig.~\ref{sfig:heating}(b).
For strong drives ($H_F\not\approx H_0$), the system initially in a thermal state for $H_0$ is strongly kicked in the first few cycles and then gets stabilized.
This early-stage dynamics (Floquet prethermalization) is well approximated by the unitary evolution by $H_F$ rather than $H_0$~\cite{Abanin2015,Kuwahara2016,Howell2019}.
The Floquet FGR describes the heating after the perthermalization.

We extract the heating rates at a fixed value $\epsilon=-0.48$ from the simulated dynamics as follows.
We find $k_0$ such that $\epsilon_{k_0-1}<-0.48<\epsilon_{k_0}$.
To extract the slope of the energy density $\epsilon_k$, we take $K$ data points $\{\epsilon_k\}_{k=k_0}^{k_0+K}$.
In this work, we have used $K=20$ to reduce the effect of fluctuations.
The insets of Fig.~\ref{sfig:heating} show the linear least squares fit for two representative $h_x=1.5$ and $5.5$.
To convert into the heating rate $d\epsilon/dt$, we divide the slope by the driving period $T$.

\end{document}